# Using Tracker as a Pedagogical Tool for Understanding Projectile Motion

Loo Kang WEE[1], Charles CHEW[2], Giam Hwee GOH[3], Samuel TAN[1], Tat Leong LEE[4]

[1] Ministry of Education, Education Technology Division, Singapore
[2] Ministry of Education, Academy of Singapore Teachers, Singapore
[3] Ministry of Education, Yishun Junior College, Singapore
[4] Ministry of Education, River Valley High School, Singapore

wee_loo_kang@moe.gov.sg, charles_chew@moe.gov.sg, goh_giam_hwee@moe.edu.sg, lee_tat_leong@moe.edu.sg, samuel_tan@moe.gov.sg

Abstract:
This paper reports the use of Tracker as a pedagogical tool in the effective learning and teaching of projectile motion in physics. When computer model building learning processes is supported and driven by video analysis data, this free Open Source Physics (OSP) tool can provide opportunities for students to engage in active inquiry-based learning. We discuss the pedagogical use of Tracker to address some common misconceptions of projectile motion by allowing students to test their hypothesis by juxtaposing their mental models against the analysis of real life videos. Initial research findings suggest that allowing learners to relate abstract physics concepts to real life through coupling computer modeling with traditional video analysis could be an innovative and effective way to learn projectile motion.
2015 Resources: http://iwant2study.org/ospsg/index.php/interactive-resources/physics/02-newtonian-mechanics/01-kinematics/174-projectile-motion

Keyword: Tracker, active learning, education, teacher professional development, e-learning, open source physics, GCE Advance Level physics
PACS: 01.40.gb 01.50.H– 01.50.ht 01.50.hv 45.50.Dd

## I. INTRODUCTION

Despite many attempts by educators to bring in hands-on activities such as real equipment [1], video analysis [2] or computer simulation [3] in the learning of projectile motion, some research studies [4-6] continue to document the misconceptions or learning difficulties encountered by students.

While the use of real life examples, such as tossing a ball to demonstrate projectile motion, is performed by students in class, it is nevertheless challenging for students to understand the mathematical equations involved in projectile motion.

Our hypothesis is that by allowing students to test their mathematical computer models against video analysis of the real motion of a projectile, students will gain a deeper understanding of the concepts and overcome the learning difficulties associated with the topic.

To test this hypothesis, a relatively new pedagogical approach called 'video modeling' [7] by Douglas Brown is used. The free software tool Tracker [8] used in this 'video modeling' can be downloaded from the Open Source Physics [9] website and has been used by authors [10] in Physics Education journal as well.

## II. INSTALLATION OF TRACKER

Tracker is a video analysis and modeling tool built on the Open Source Physics (OSP) Java framework. Though it is possible to run from the Web start or a 3.9 Mb Tracker_461.jar file, we recommend using the respective installers found at http://www.cabrillo.edu/~dbrown/Tracker/, especially to enable the Xuggle video engine [11] that can decode most video file formats. Installers for Tracker version 4.62 installers are available in Windows, Mac OS X as well as Linux operating systems.

## III. VIDEO ANALYSIS OF PROJECTILE MOTION

In an ideal projectile motion, equations (1) and (2) represent the mathematical equations of velocity and displacement in the $x$ and $y$ direction respectively.

$$\begin{pmatrix} v_x \\ v_y \end{pmatrix} = \begin{pmatrix} u_x \\ u_y \end{pmatrix} + \begin{pmatrix} g_x \\ g_y \end{pmatrix} t \quad (1)$$

$$\begin{pmatrix} x \\ y \end{pmatrix} = \begin{pmatrix} x_0 \\ y_0 \end{pmatrix} + \begin{pmatrix} u_x \\ u_y \end{pmatrix} t + \frac{1}{2} \begin{pmatrix} g_x \\ g_y \end{pmatrix} t^2 \quad (2)$$

After using Tracker to either manually or automatically track the projectile's trajectory, the data tool can be used to analyze motion and allows the determination of the numerical values of quantities such as acceleration, $g_x$ and $g_y$ and initial velocity $u_x$ and $u_y$, in the $x$ and $y$ directions respectively. This can be done by comparing the coefficients of the respective linear fit and parabolic fit as in equations (3) and (4) with equation (1) and (2) later on.

$$\begin{pmatrix} v_x \\ v_y \end{pmatrix} = \begin{pmatrix} b_x \\ b_y \end{pmatrix} + \begin{pmatrix} a_x \\ a_y \end{pmatrix} t \quad (3)$$

$$\begin{pmatrix} x \\ y \end{pmatrix} = \begin{pmatrix} c_x \\ c_y \end{pmatrix} + \begin{pmatrix} b_x \\ b_y \end{pmatrix} t + \frac{1}{2} \begin{pmatrix} a_x \\ a_y \end{pmatrix} t^2 \quad (4)$$

The video on the Tracker installation called BallTossOut.mov (Figure 1) is used for subsequent discussions in this paper. Readers who need step by step help in video analysis may refer to this YouTube video [12]. There is a slight difference between the current version 4.62 and 3.1 but the YouTube video is still relevant.

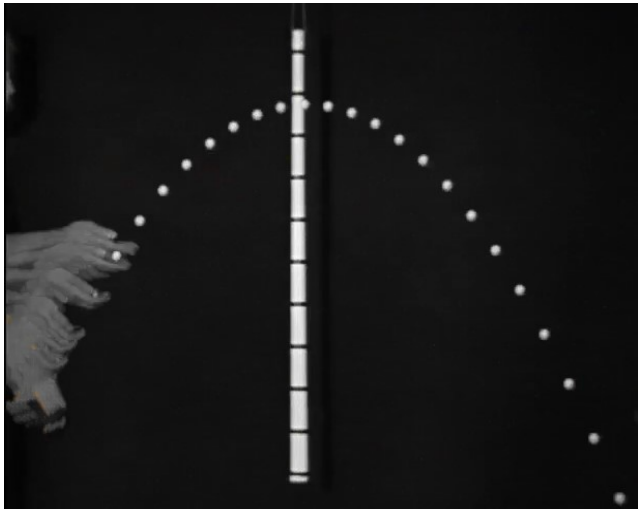

Figure 1. World view of a projectile motion as in the video 'BallTossOut.mov' shown with 0 fade using the ghost filter.

As an example, Figure 2 shows a typical Tracker's video analysis of a $y$ vs $t$ graph. After choosing the parabola fit of equation (4y), the parameters $a, b$ and $c$ are determined by the Tracker's data tool as shown by the "Fit Equation" and compared to coefficients in equation (4y) and subsequently to equation (2y).

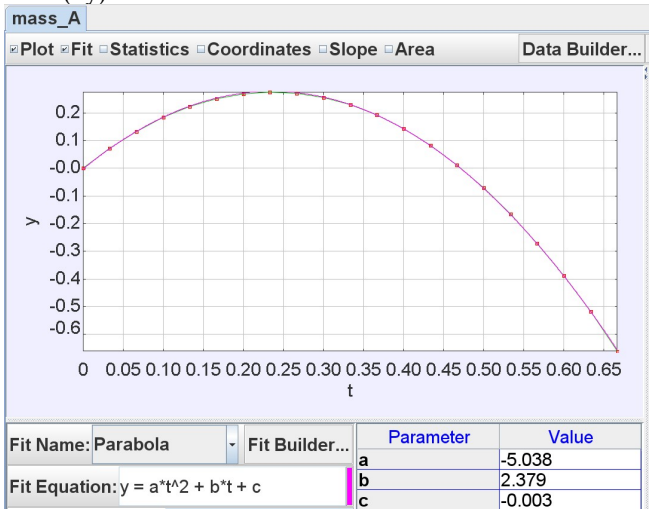

Figure 2. Data tool display of mass_A $y$ versus $t$ view where parabola fit equation of $y = a^2*t + b*t + c$ is used with parameters $a = -5.038$, $b = 2.379$ and $c = -0.003$. By comparing with equation (4y), it is determine that $g_y = -10.08$ ms$^{-2}$ and $u_y = 2.379$ ms$^{-1}$ respectively.

Thus,
$a$ (parameter in Tracker's Data Tool) = $a_y$ (equation 4y) = $-5.038$ (value in Tracker's Data Tool),
$b = b_y = -2.379$ and
$c = c_y = -0.003$.
By comparing coefficients with the equation (2y), the students can infer the values of:

$\frac{1}{2} g_y = -5.038$, thus $g_y = -10.08$ ms$^{-2}$,
$u_y = 2.379$ ms$^{-1}$ and
$y_0 = -0.003$ m respectively.

The value $a_y = -10.08$ ms$^{-2}$ can be interpreted to be approximately equal in value to the gravitational acceleration constant at the surface of Earth of $-9.81$ ms$^{-2}$ (3 significant figures adopted by Advanced Level General Certificate of Education GCE) and $-10$ ms$^{-2}$ (2 significant figures required by GCE Ordinary Level).

Similarly, video analysis can be conducted on the $x$ vs $t$ graph (Figure 3) to arrive at the values of:
$u_x = 1.733$ ms$^{-1}$ and
$x_0 = -0.005$ m *respectively.*

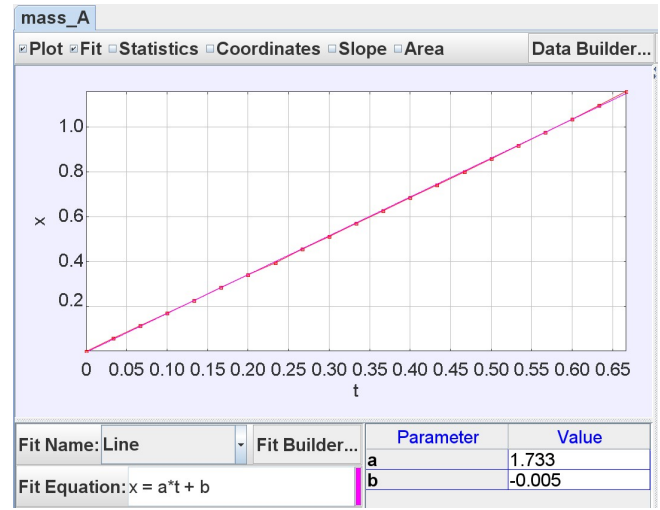

Figure 3. Data tool display real data mass_A of the $x$ versus $t$ view where a line fit equation of $x = a*t + b$ is used with parameter $a = 1.733$ and $b = -0.005$ as determined using Tracker. By comparing with equation (4x), it is determine that $v_x = 1.733$ ms$^{-1}$.

A good teaching point that surfaces is the appreciation of a possible systematic error from the calibration stick length positioning, and random errors from the measurement or selection of the analysis points, especially when manually tracking the path of the motion.

This active inquiry activity typifies the actions of scientists [9,13], and is well suited for students to construct their own understanding of the physics of motion through a real world video, without referring to the authoritative sources of knowledge such as teachers and books.

Only after this determination of the variables through video analysis, can we tap on the pedagogical advantage of video modeling to allow students to test out their understanding of projectile motion and also to address their misconceptions in the process of model construction.

IV. VIDEO MODELING – DYNAMIC PARTICLE MODEL AS A PEDAGOGICAL TOOL THAT CAN ADDRESS MISCONCEPTIONS

Novice students who develop personal "theories of motion" [14] by generalizing the ideas they acquire from observation of objects in everyday situations [4] and non-projectile motion, such as propelled rockets seen in entertainment media, can harbor naive "impetus theory" [5] of motion. Thus, we suggest using Tracker's dynamic particle instead of the other kinetic particle, because the dynamic model easily represents gravity and drag forces affecting the motion. This dynamic model when utilizing numerical data values determined from earlier III video analysis to test their "impetus theory" models juxtaposed against a real video of projectile motion, can address some misconceptions as discussed later in IV.A, IV.B and IV.C.

Again, readers may refer to this other YouTube video [15] that shows the process of building a dynamic particle model on a projectile motion video useful.

### A. Absence of a x-direction force

According to naive theory of impetus [16],

"When a mover sets a body in motion he implants into it a certain impetus, a certain force enabling a body to move in the direction in which the mover starts in".

Our research suggests that some students do think of an object in motion as having external force acting on it, for it to continue in the $x$ direction. Although this can be verbally rationalized with students using Newton's First Law of motion that since there is constant $x$ velocity, the resultant force in $x$ direction has to be zero. However, without an opportunity to refute their incorrect mental model, this "talk" [17] alone may not be effective in helping students to advance their conceptual understanding.

Using Tracker, we guide students to test out their mental model by proposing a dynamic particle model that has a non-zero value in the force in x-direction, $f_x$, inferring from equation (3x) and (4x) when $a_x = 0$ ms$^{-2}$ and mass = 1 kg.

We suggest an activity where students key in values for the initial velocity $v_x$ in the dynamic model and observe the real data (red) versus the constant $v_x$ model (pink). This allows them to make sense for themselves that instantaneous velocity is (in this case) approximately equal to 1.733 ms$^{-1}$ at all times of the projectile motion as in Figure 4.

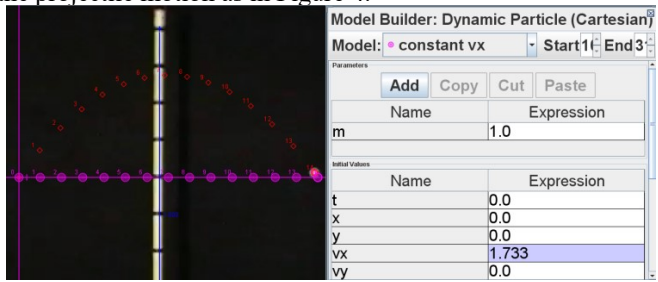

Figure 4. World view of a projectile motion with real data (red) versus the constant $v_x$ model (pink) on the left and the model builder of $v_x = 1.733$ ms$^{-1}$ on the right. Notice the pink model of constant $v_x = 1.733$ ms$^{-1}$ is a vertically downward projection of the red real motion.

Similarly, by keying in non-zero values for $f_x$ ($f_x \neq 0$ N) when mass of projectile $m = 1$ kg, students can observe paths similar to Figure 5 for the case of $f_x = 10$ N and compare the real data (red) versus the $f_x = 10$ N model (teal). Students can verify that the vertically projected downward 'shadow' of the real data does not coincide with the $f_x = 10$ N model (teal) and thus, will continue to propose different values until they are satisfied with their own model of $f_x = 0$ N (pink) as in earlier Figure 4.

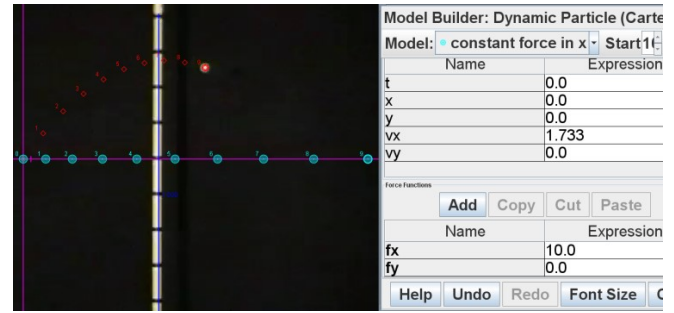

Figure 5. World view of a projectile motion with real data (red) versus the $f_x = 10$ N model (teal) on the left and the model builder values of $v_x = 1.733$ ms$^{-1}$ and $f_x = 10$ N for this incorrect model on the right. The evidence of the incorrect model against the real data compels students to rethink their assumptions on non-zero force in x-direction

### B. The y direction acceleration is constant and has a negative sign

Many students espouse naive impetus theory such as the need for an upward force in projectile motion, similar to the x-direction as in the case of absence of a x-direction force mentioned earlier in *A*.

Students are guided to key in values for positive values for $f_y$ and infer that their naïve impetus theory cannot be validated with the real video.

As the computer axis adopts the conventional Cartesian coordinate system, this allows students to appreciate the negative sign of $g_y = -10.08$ ms$^{-2}$ easily as a positive $f_y$ results in a parabolic upwards path (Figure 6).

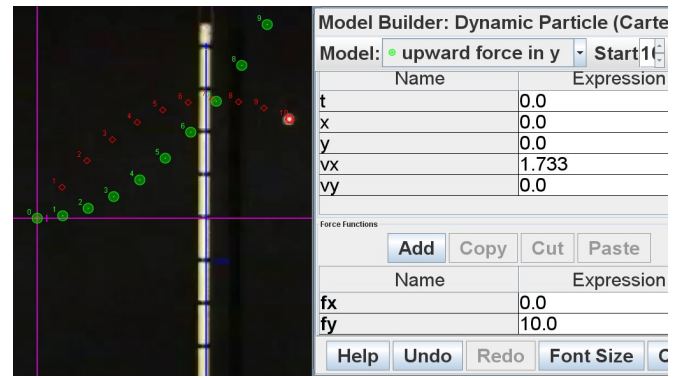

Figure 6. World view of a projectile motion with real data (red) versus the $f_y = 10$ N model (green) on the left and the model builder values of $v_x = 1.733$ ms$^{-1}$ and $f_y = 10$ N for this incorrect model on the right. The evidence of the incorrect model against the real data compels students to rethink their assumptions on positive force in y-direction.

By incrementally improving the computer model, students can develop deeper understanding of the consequence of changing $f_y = -10$ N (Figure 7).

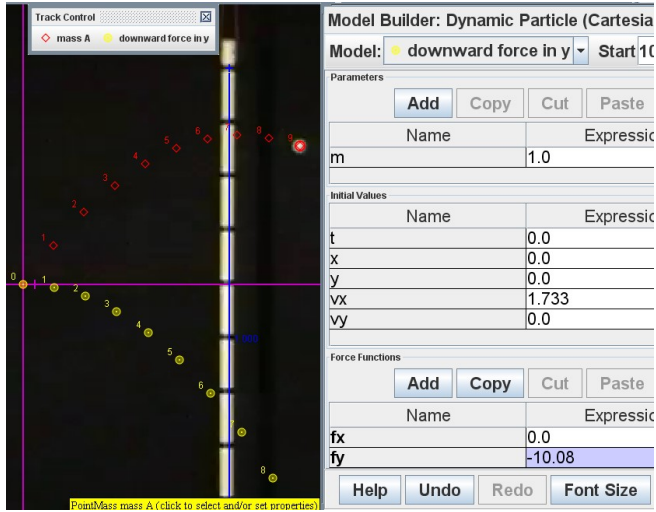

Figure 7.  World view of a projectile motion with real data (red) versus the $f_y = -10.08$ N model (yellow) on the left and the model builder values of $v_x = 1.733$ ms$^{-1}$ and $f_y = -10.08$ N for this more suitable model on the right. This model provides evidence of possibly a correct downward direction of the force.

Figure 8 shows how a complete model will look like with the inclusion of initial velocity in $y$ direction, $v_y = 2.379$ ms$^{-1}$.

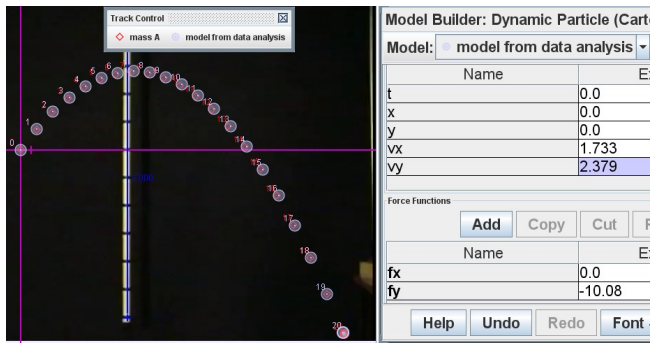

Figure 8.  World view of a projectile motion with real data (red) versus the correct model of $v_x = 1.733$ ms$^{-1}$, $v_y = 2.379$ ms$^{-1}$ and $f_y = -10.08$ N (light blue) by a data driven modeling process instead of trial and error can deepen learning.

We recommend a data driven modeling process to deepen learning, i.e. students use initial values obtained from the analysis process as in III instead of a completely trial and error approach.

### C. Absence of air resistance in low velocity projectile motion

Expert students can be challenged to extend their own learning to model air resistance by comparing the trajectories of the dynamic particle model and that of the real video, and being convinced that the video's projectile motion cannot be realistically assumed to be a motion with significant air resistance Figure 9. The model for air resistance may be expressed as in equation (5) with $f_{drag} = k*v$ (assuming a simple linear relationship between drag force and velocity it is traveling at).

$$\begin{pmatrix} f_{x\,drag} \\ f_{y\,drag} \end{pmatrix} = k \begin{pmatrix} v_x \\ v_y \end{pmatrix} \quad (5)$$

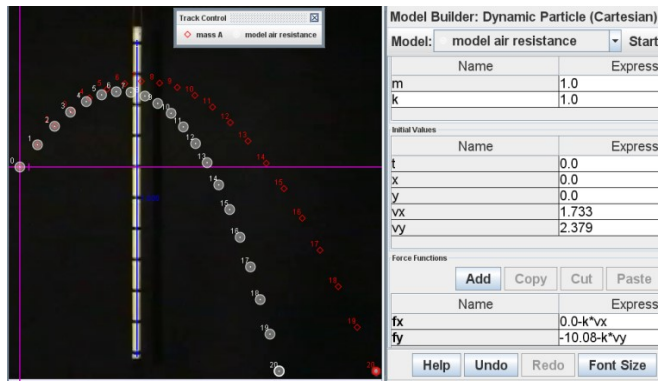

Figure 9.  World view of a projectile motion with real data (red) versus the air drag model by $F_{drag} = k*v$ by inserting $f_x = 0 - k*v_x$ and $f_y = -10.08 - k*v_y$ (white) where $k = 1$ on the left with the model builder values on the right.

Only when the values of $k$ are very close to zero do the trajectories of the model and the real video juxtapose the closest. This again gives real evidence to conclude that the tossing out of a ball in projectile motions at low velocities are well approximated by the theoretical equations (1) and (2).

### V. STUDENTS' REFLECTION ON TRACKER LESSON

To give some themes into the conditions and processes during the laboratory lessons, the following are some excerpts from the qualitative survey results and informal interviews with the students. Words in brackets [ ] are added to improve the readability of the qualitative interviews.

*1) Active learning can be fun*

"Very good. Improve our understanding on physics, as we are able to see the real life example in a fun and interesting way."

"I liked doing the experiment with those interesting instruments and software.  I learnt a lot in a fun way."

"I think it is amazing how the comp[uter] knows how to do so many kinematics stuff, its kind of fun doing practical using this."

*2) Tracker can support inquiry learning and thinking like real scientists*

"It prepares us for similar activities we have to do in [the] future, either in university or our careers in the science industry. It is also interesting to learn how an actual process (visual form) of motion-tracking is like, instead of drawing graphs based on a worded situation."

"Get to learn things by ourselves, not spoon-feed by teachers. It is interactive; the visual is much better for learning than all the words in the tutorials."

"It was refreshing....Through this video Tracker lesson, we learned that there's such a programme for the scientist :)"

"Normally I thought those theories and formulae don't work perfectly in real life. The programme shows they actually work."

### 3) Overcoming initial difficulties using Tracker

"System [Tracker] may be difficult to use at first but once you can get use to it, it is quite convenient and very fast"

## VI. Teachers' reflection on using the Dynamic PArticle Model

We would like to share two significant reflections by the ten teachers in the using of this dynamic particle model:

1. The mathematical and graphical understanding of the motion in the vertical and horizontal directions improves as the students deduce the kinematics equations and mathematical representations during the activity.

2. Video modeling pedagogy is suitable for active and deep learning because the students can be said to be predicting by keying certain values, observing by comparing the real data with the current proposed model, and explaining [18] their choice of values and making sense of the video analysis data. Even with incorrect models proposed, the results from the world view and associated multiple representational views [19] in various scientific plots can allow the facilitation of data driven social discussions among students and teachers.

## VII. Conclusion

The relative ease of installation and use of Tracker to conduct learner-centered in-depth video analysis with reference to the theoretical physics model of an ideal projectile motion is discussed in this paper. The values deduced from video analysis are consistent with real world data of gravitational acceleration on surface of Earth.

More importantly, the video modeling especially when driven with data from video analysis, allow students to discover using evidences and incomplete models proposed by themselves, to incrementally and iteratively improve and self-invent a better model to predict and explain the projectile motion. This has lead to surprisingly pleasant "Eureka" experiences for our students as they connect abstract concepts and formulae with real world examples thus deepening learning.

### Acknowledgment

We wish to acknowledge the passionate contributions of Douglas Brown, Wolfgang Christian, Mario Belloni, Anne Cox, Francisco Esquembre, Harvey Gould, Bill Junkin, Aaron Titus and Jan Tobochnik for their creation of Tracker video analysis and modeling tool.

### Author

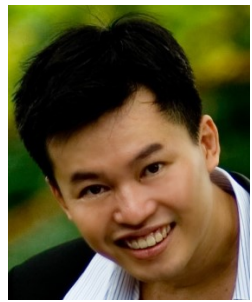

Loo Kang WEE is currently an educational technology officer at the Ministry of Education, Singapore and a PhD candidate at the National Institute of Education, Singapore. He was a junior college physics lecturer and his research interest is in Open Source Physics tools like Easy Java Simulation for designing computer models and use of Tracker.

Dr Charles CHEW is currently a Principal Master Teacher (Physics) with the Academy of Singapore Teachers. He has a wide range of

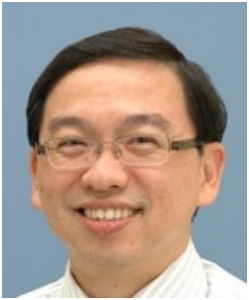 teaching experiences and mentors many teachers in Singapore. He is an EXCO member of the Educational Research Association of Singapore (ERAS) and is active in research to strengthen theory-practice nexus for effective teaching and meaningful learning.

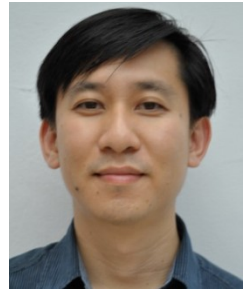 Samuel TAN is an educational technology officer at the Ministry of Education, Singapore. He designs ICT-enabled learning resources for Science and develops ICT pedagogical frameworks for implementation at schools.

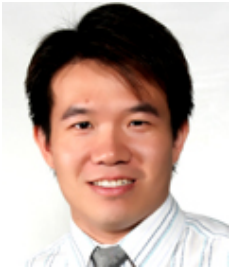 Giam Hwee GOH is currently the Head of Science Department in Yishun Junior College, Singapore. He teaches Physics to both year 1 and 2 students at the college and advocates inquiry-based science teaching and learning through effective and efficient means.

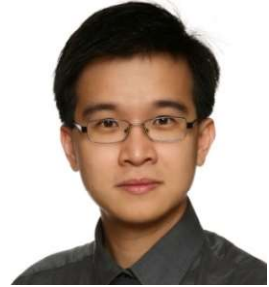 Tat Leong LEE is currently the Head of Department for Education Technology in River Valley High School, Singapore. He is a high school Physics teacher, with 10 years of teaching experience. He has been using Open Source Physics (OSP) tools as early as 2006 (Tracker and Easy Java Simulations).